# Albert Einstein as the Father of Solid State Physics

Manuel Cardona

*Max-Planck-Institut für Festkörperforschung, Heisenbergstrasse 1, 70569 Stuttgart, Germany*

(Dated: August 2005)

Einstein is usually revered as the father of special and general relativity. In this article, I shall demonstrate that he is also the father of Solid State Physics, or even his broader version which has become known as Condensed Matter Physics (including liquids). His 1907 article on the specific heat of solids introduces, for the first time, the effect of lattice vibrations on the thermodynamic properties of crystals, in particular the specific heat. His 1905 article on the photoelectric effect and photoluminescence opened the fields of photoelectron spectroscopy and luminescence spectroscopy. Other important achievements include Bose-Einstein condensation and the Einstein relation between diffusion coefficient and mobility. In this article I shall discuss Einstein's papers relevant to this topic and their impact on modern day condensed matter physics.

## 1. 1900-1904

### 1.1 Einstein's first publication

Albert Einstein started his career as a scientific author on Dec. 13, 1900 when he submitted an article to the *Annalen der Physik*, at that time probably the most prestigious and oldest physics journal. He was then 21 years old. The author's by-line lists him simply as "Albert Einstein", Zürich, without mentioning any affiliation. The article was rapidly accepted and it appeared the following year.[1] He had come across, while searching the literature, a collection of data on the surface energy of a number (41) of complex organic liquids containing several of the following atoms: C, O, H, Cl, Br, and I (e.g. benzylchloride: $C_7H_5OCl$). He proceeded to develop a phenomenological theory for fitting all 41 surface energies of these liquids on the basis of a small number of adjustable parameters (6) associated with the six atoms present. He reasoned as follows:

The energy of the liquid with or without a surface is obtained by summing the contributions of all possible pairs of molecules which interact with each other through a molecular pair potential. The pairs to be summed will be different inside the liquid and at the surface. The difference constitutes the surface energy. He now assumes that the pair energy is given by a universal function of the intermolecular distance, with a prefactor which is the sum of corresponding numbers characteristic of the atoms involved, six different ones for the cases under consideration. In this manner, by adjusting the 6 atomic coefficients, he obtained a rather good fit to the surface energies of all the liquids under consideration.

This article, like most of his publications prior to 1933, was written in Einstein's very elegant German [I found two articles in English published before his forced emigration in 1933. One appeared in *Nature* in 1921, the other in the *Physical Review* in 1931]. It has been cited 33 times and it is still being cited to date[2]. Its subject falls into the category of physical chemistry and exemplifies the fact, found not only in his first paper, that Einstein would tackle anything that he felt he could make an impact on, regardless of how pedestrian. Later on, we find that he works on simple problems before or after tackling the most sublime ones for which he is known. His first publication could already be considered to be in the realm of Condensed Matter Physics (liquids). It contains what is probably the first example of the use of pair potentials in condensed matter physics.

In 1902 Einstein submitted his first and second articles as a PhD thesis to the University of Zurich. The reviewer, Prof. Kleiner, rejected them.

**1.2 Einstein's second publication**

Einstein's second article was submitted to and appeared in the "Annalen" in 1902.[3] It also corresponds to the field of physical chemistry. In contrast to his first paper, where he develops a semiempirical theory to interpret extant experimental data, in this article he develops a theory, based on thermodynamics, which should have been helpful to interpret a large number of experiments concerning contact potentials between metals and their fully dissociated salts in solution. He uses the method of Ref.1 (intermolecular forces) to calculate the effect of the solvent on the contact potentials. This rather long (16 pages) and comprehensive article has not received much attention, having been cited only 7 times. Einstein himself seems to have had a premonition of the reduced interest this article may trigger. He closes it with a statement, again in rather flowery but poignant German, expressing his feelings. The closest I can come up with in English is:

*In conclusion, I feel the urge to apologize for the fact that I have only developed in this article a clumsy plan for a painstaking investigation without having contributed to its experimental solution; I am simply not in a position to do it. However, this work will have reached its aims if it encourages some scientist to tackle the problem of molecular forces with the method I have suggested.*

Present day physics editors would most likely not allow such tirades involving a combination of hard core science and personal feelings. Such statements are not unusual in other publications of Einstein and give us a glimpse into his psyche and/or sense of humor that we miss in the current literature.[4]

**1.3 Einstein's three additional publications in the "Annalen" before the of**
   *annus mirabilis*

Einstein submitted and published three articles in the "Annalen" during the years 1902-1904. They dealt with kinetic theory, the foundations of thermodynamics and the general molecular theory of heat.[5 6 7] These papers resulted from his attempts at teaching himself the disciplines of thermodynamics, kinetic theory and statistical mechanics. His knowledge of the work of Boltzmann was rather fragmentary and he does not seem to have been aware, at that time, of the treatise of J.W. Gibbs (Elementary Principles of Statistical Mechanics, 1902). So, he rediscovered much of the material already existing on these subjects. In accepting these papers, the responsible editor of the "Annalen" did not seem to be aware of those works either. Nevertheless, the published articles by Einstein reveal his unique way of arriving to the basic concepts of thermodynamics and kinetic theory, in particular entropy and the second principle. Following the tempers of the times (and Einstein's) these papers contain very few citations, only to Boltzmann and to Einstein himself. I found particularly interesting the treatment of energy fluctuations in a system in thermal equilibrium with a reservoir (Ref.6), which he masterfully applied in many subsequent papers.[4,8]

In autobiographical notes published in 1949[9] Einstein wrote *"Unacquainted with the investigations of Boltzmann and Gibbs, which had dealt exhaustively with the subject, I developed statistical mechanics and the molecular-kinetic theory of thermodynamics..."*.
In 1910 Einstein had already written that had he known of Gibbs's book he would not have published Refs. 4-6.[10]

## 1. ANNUS MIRABILIS: 1905

Apparently in his "spare time", while working at the Swiss patent office in Bern, Einstein wrote five revolutionary papers and submitted them to the "Annalen". Except for one, which he withheld for a few months, in order to incorporate in it the most recent experimental data[11], they were quickly accepted and published. Reference 11 was submitted to the University of Zurich as a PhD thesis. This time Prof. Kleiner approved it and Einstein became a doctor.

The topics of the five famous papers submitted in 1905 to the "Annalen" and the corresponding references are give below:

1. The quantum of light, the photoelectric effect and photoluminescence[12]
2. The theory of Brownian motion[13]
3. Special Relativity[14]
4. The dependence of the inertial mass on energy[15]
5. Determination of the size of a molecule and Avogadro's number[11]

### 2.1 The quantum of light, the photoelectric effect and photoluminescence

This work was published in Ref. 12. It is actually the work that was mentioned in the citation of the 1921 Nobel prize (...for your work on theoretical physics and, in particular, for your discovery of the law of the photoelectric effect). This citation already appeared in the notification from the Nobel Foundation he received by cable on November 10, 1922. The telegram mentioned explicitly that his work on the theory of relativity was not considered for the award (see Ref. 9, p.503). The award of the 1921 prize had been deferred, probably because of pressure to honor the theory of relativity, a possibility which was not acceptable to some conservative members of the Nobel committee. Once the proposal of the photoelectric effect was on the table, objections vanished and Einstein was belatedly awarded the 1921 Prize in 1922.[16]

The possible lack of courage, or understanding of the revolutionary relativity theory, reflected by the actions of the committee, has puzzled historians and physicists for many years. More recently, however, commentators have reached the conclusion that Ref. 12 was indeed even more revolutionary than the special relativity article (Ref. 14). The mathematical underpinnings of the latter had been largely worked out by Lorentz and Poincaré. Einstein provided its philosophical underpinnings and derived the famous law of equivalence of mass and energy:[15]

$$E=mc^2 \qquad (1)$$

In Ref. 12, however, Einstein introduces for the first time the quantization of the electromagnetic (light) energy, something that was not explicitly done by Planck when developing his famous law of the black body radiation:

$$\rho_\nu = \frac{8\pi\nu^2 h\nu}{c^3} \frac{1}{e^{h\nu/kT}-1} \qquad (2)$$

While Planck assumed "as an act of desperation" that the electromagnetic energy was distributed in finite amounts ($E_\nu=h\nu$) among a large number of fictitious harmonic oscillators, Einstein considered the high frequency limit of Eq. (2), the so-called Wien's law, and derived the corresponding entropy. He then showed that this entropy equals that of an ensemble of non-interacting point-like particles with energy $E_\nu=h\nu$. While recognizing that ondulatory phenomena impose wave character to light, he realized that a number of contemporary experiments (e.g. photoemission) could only be explained by assuming that light consists of particles whose energy is proportional to the nominal frequency of the radiation:

$$E_\nu=h\nu \qquad (3)$$

These particles had to wait 20 years before being given the name of *photons*.[17] The wave-particle duality of "photons" introduced in Ref. 12 is viewed by many as an even more revolutionary step that the special theory of relativity, a fact which, in retrospect, justifies the citation which accompanied Einstein's Nobel Prize. Be it as it may, Refs. 12 and 14 firmly establish Einstein as the father of the two main tenets which revolutionized physics in the early 20$^{th}$ century: relativity and energy quantization. As we shall see below, both these tenets were to have a profound influence in condensed matter physics.

Typical of Einstein, he searched the experimental world for facts that would support his theory of light quantization. Reference 12 contains a "large" number of references, unusual for an Einstein publication and also for the customs of the times: two to Planck, three to Lenard and one to Stark, plus the mention of a few other colleagues in the text (Boltzmann, Drude, Wien).[18]

Lenard's experiments had shown that electrons were only emitted from metals (the photoelectric effect) when the frequency of the impinging light was larger than a given value, which was *independent of the light intensity but may vary from metal to metal*. This simple experimental fact cannot be explained on the basis of the wave nature of light. Its explanation is straightforward under the corpuscular assumption: the energy of each light corpuscle (photon) $E_v = hv$ must be larger than the minimum energy $I$ it takes to remove an electron from the metal, the so-called work function of the metal. The maximum energy $E_e$ of a photoemitted electron must be positive and given by:

$$E_e = hv - I . \qquad (4)$$

For photoemission to occur $hv \geq I$. For $hv > I$ the maximum energy of the photoemitted electrons increase linearly with $v$. The photoelectron current depends on the light intensity but not its energy distribution. Equation (4) is the basis of a large number of spectroscopic techniques nowadays essential for the investigation of solids, in particular for the highly topical high $T_c$ superconductors.[19]

Before moving to the next phenomenon dealt with in Ref. 12, I would like to give two examples of the current use of photoemission thresholds, as represented by Eq. (4). This equation may be interpreted as meaning that for photon frequencies such that $hv < I$ no electrons whatsoever will be emitted, i.e. that the emitted current will show a sharp step for $hv = I$. In spectroscopy, sharp steps seldom occur: they are usually smeared out either by the experimental resolution, by impurities and disorder, or by thermal fluctuations. The latter are represented in metals by the Fermi-Dirac distribution function (1926), a fact which was unknown to Einstein in 1905. Figure 1 shows the photoelectric yield of gold (in a logarithmic scale) vs. the energy of the exciting photons (in eV, measured at 300 K). Below the so-called Fermi energy $E_F$, which corresponds to the work function, the yield plummets rather steeply, falling down by 2 orders of magnitude for every 0.1 eV. The thick red line represents a calculation based on the convolution of an infinitely steep edge ($T = 0$) and the Fermi-Dirac function for 300 K. Figure 1 confirms that the corpuscular theory of light, coupled to the Fermi-Dirac distribution, represents rather well the photoemission threshold of gold. The threshold spectrometer used for the work[20] in Fig. 1 has a very large dynamical range (ten decades).

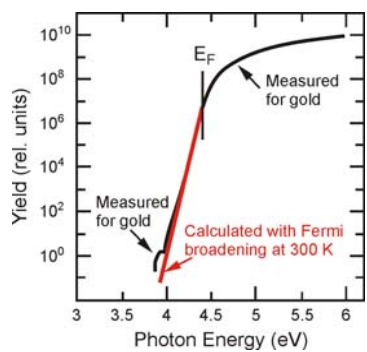

**Fig. 1** Photoelectric yield near the threshold for electron emission of gold. The broadening below $E_F$, encompassing eight decades of yield, is due to the thermal smearing of the Fermi distribution at 300K. From ref. 20.

In its construction, currents produced by spurious electrons must be carefully avoided. This type of instrument is used nowadays to investigate impurity, surface and defect states within the gap of semiconductors.[21] But perhaps the most spectacular application of photoelectron spectroscopy is the so-called angular resolved photoemission spectroscopy (ARPES). In this technique, the electrons escaping along a certain space direction are measured for several directions of momentum space. A threshold corresponding to the Fermi surface is seen in the spectra vs. electron energy (Fig.2). The transition from a normal metal to a superconductor is accompanied by the opening of an energy gap around the Fermi surface and the concomitant shift of the photoemission threshold. Limited resolution of ARPES instruments hinders their application to conventional superconductors. It has been, however, very useful for the investigation of high $T_c$ materials because of their larger gap. Figure 2 indicates that the superconducting gap $Bi_2Sr_2Ca\,Cu_2O_y$ is strongly anisotropic, a fact which seems to be crucial for understanding these materials.[19]

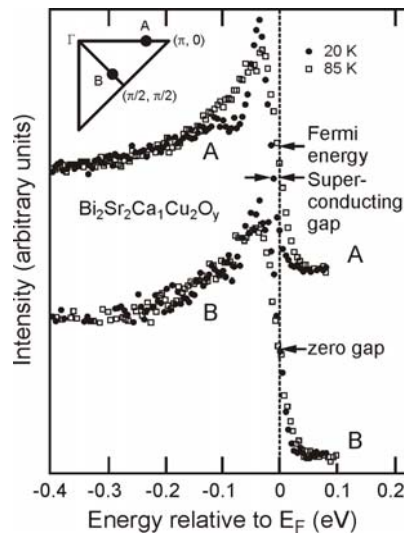

**Fig. 2** Photoelectron spectra of a high $T_c$ superconductor ($Bi_2\,Sr_2Ca_1\,Cu_2O_\delta$) obtained with a high resolution angle resolved spectrometer above $T_c$ (85 K) and below $T_c$ (20K). The sharp thresholds correspond to the Fermi energy. The shift of this threshold from 85K to 20K represents the opening of the superconducting gap. Note that this shift does not appear in the lower curves, a fact that signals the anisotropy of the superconducting gap, one of the most striking properties of these materials. From Ref. 19.

Einstein's publication on the quantum nature of light also discusses two other experimental results. The first one has to do with the light emitted by a solid upon illumination, the so-called photoluminescence. It was known at that time that the emitted light had a frequency somewhat smaller than that of the exciting radiation, independent of the strength of the latter. Einstein realized that this so-called Stokes rule follows immediately from the corpuscular theory of light: a photon with energy $h\nu$ impinging on matter, will produce an elementary excitation with energy equal to or smaller than $h\nu$. Conversion of this excitation energy into light will generate photons of energy (i.e. frequency) lower than that of the incident photon. Hence, the Stokes rule follows. Einstein concludes the corresponding section of the paper by pointing out that violations of the Stokes rule may occur (today we speak of anti-Stokes radiation). He mentions two possible mechanisms:

1. Thermal excitation at sufficiently high temperature which will provide a higher excitation energy than that of the incident photons.

2. Nonlinear optical effects: the outgoing photon may result from the merger of two or more excitations if high intensity light is impingent on the material.

It is most remarkable that Einstein would have thought about nonlinear optical processes. It took more than 50 years, and the advent of lasers, to be able to effect such nonlinear processes which are now standard manipulations in nonlinear optics.

Finally, Einstein mentions the ionization of gases by ultraviolet radiation, which is also easily accounted for by the corpuscular theory. Here he uses again experimental data by Lenard and also experiments on ionization by applied electric fields performed by Stark (i.e. Aryan Physics).

## 2.2 The theory of Brownian motion

The manuscript on the corpuscular nature of light has an entry at the end signifying that in was finished on March 17 1905. Four lines below it there is a byline saying that the manuscript was received by the editors of the *Annalen* on March 18, 1905. Not a clue as to how this was possible. It took Einstein 6 weeks to finish his doctoral thesis based, as already mentioned, on Ref. 11.[22] On May 11, 1905 his manuscript on Brownian motion was received at the editorial office of the *Annalen*.

Einstein meticulously avoids calling the work in Ref. 13 Brownian motion. He writes, however "*possibly this motion is identical with the so-called Brownian motion, however the information available to me is so imprecise that I cannot make a judgment*" Again, I doubt that present day editors would be willing to print such a statement.

In this article, with ~1520 citations one of his most highly cited ones, Einstein derives an expression for the average distance traveled by a suspended particle under the influence of collisions with the solvent molecules in a time *t*:

$$<r^2>^{1/2} = (6D\ t)^{1/2} \qquad (5)$$

where *D* is the diffusion coefficient for the suspended particles, for which he derives the famous expression:

$$D = \frac{RT}{N} \cdot \frac{1}{6\pi\eta a} , \qquad (6)$$

where $R$ is the gas constant, $\eta$ the viscosity of the solvent, $a$ the average radius of the suspended particles and $N$ Avogadro's number, a number which seems to have fascinated Einstein as it embodied the corpuscular theory of matter.

He then proceeds to estimate the diffusion length $<r^2>^{1/2} = 6$ μm for $t =1$ min. at T=17°C using the viscosity of water and the value N = 6 x$10^{23}$ mol$^{-1}$ obtained from the kinetic theory of gases.

In the abstract he mentions that agreement of his prediction of the diffusion length with experiment would be a strong argument in favor of the corpuscular theory of heat. Conversely, if experiments do not confirm his predictions, it would be a strong argument against such theory. He concludes this article with a typical Einstein statement:
*Let us hope that soon a researcher will decide among the questions presented in this paper, which are very important for the theory of heat.*

A later article[23], submitted again to the *Annalen* in Dec. 2005, starts by mentioning that the phenomenon treated in Ref. 21 was indeed the so-called Brownian motion. An article by Gouy[24], in which the random motion was attributed to the thermal motion of the fluid, had been brought to Einstein's attention by a colleague from Jena (C. Siedentopf). Having thus exculpated himself of omitting to cite Gouy's work, he takes up the Brownian motion again and calculates the angular fluctuations of a spheroidal particle in suspension as induced by the thermal agitation. He then points out that Eq. (5), and the equivalent one for angular fluctuations, is only valid at sufficiently large times. He then estimates the minimum times at which it should remain valid, giving the value of $10^{-7}$ sec. for typical particles of 1 μm diameter.

2.3 **A new determination of the molecular dimensions**[11]

As already mentioned, Einstein seems to have had a fixation with Avogadro's number *N*. He suggested ~ 8 different methods for its determination from experimental data. In Ref. 11 he presents a method to determine both, *N* and the radius *a* of a molecule. This work was submitted as a doctoral dissertation and accepted by the University of Zürich. With ~1622 citations, it is the most cited of Einstein's papers with the exception of his rather late (1935) paper on the incompleteness of quantum mechanics, the so-called EPR paradox.[25]

Reference 11 describes a very ingenious technique to simultaneously determine *N* and the molecular radius *a* from experimental data. For this purpose Einstein uses measurements of the increase in viscosity effected by dissolving sugar into water. By means of a non-trivial hydrodynamic calculation he finds for the viscosity η* of such a solution:

$$\eta^* = \eta \, [ \, 1 + (5/2) \, \varphi \, ] \qquad (7)$$

where $\varphi$ represents the fraction of the solution volume occupied by the molecules, taking into account that in solution a layer or more of water is attached to the molecule (one may speculate how he figured this out without having been exposed to much chemistry!). The original article does not contain the factor (5/2). Because of an error, which Einstein admits, in the rather complex hydrodynamic calculation.[26] Adding the 993 citations to this erratum to those of Ref. 11, we find 2615 citations, now even higher than those received by the EPR article.[25] The determination of $\varphi$ from the experimentally observed increase in viscosity using Eq. 7 provides a relationship between Avogadro's $N$ and the molecular radius. A second relation is needed in order to determine $N$ and $a$ separately. For this purpose, Einstein used the viscosity of the suspended molecules as given in Eq. 6.

From the point of view of the solid state physicist, Eq. 6 is rather important. The viscosity $\eta$ represents thermal losses which take place when the solute moves in the solvent, i.e. the inverse of the mobility µ of the molecules in the solvent when propelled by an external force. Equation 6 can thus be written in the following way, more familiar to semiconductor physicists:

$$D = \mu T R / N e ,  \qquad (8)$$

which is the famous Einstein relation between diffusion coefficients of carriers and their mobility in semiconductors, governing the diffusion of carriers in transistors and other devices. Equation 6, and correspondingly Eq. 8, can be viewed as the first expression in the literature of the rather important fluctuation-dissipation theorem, the diffusion coefficient $D$ describing fluctuations and the viscosity η describing dissipation.

## 2.4 **The theory of special relativity**[14,15]

It is often believed that relativity is of little importance in the fields of chemistry and condensed matter physics[27]. This is not correct: the motion of electrons in solids can reach, in certain cases, velocities close to that of light. Neglecting relativistic effects in calculations can lead to serious errors. There is copious recent literature illustrating the incorrectness of Dirac's statement.[28]

Special relativity is also the basis of important experimental techniques such as synchrotron radiation.[29] About 40 synchrotrons are now available world wide dedicated to spectroscopic investigations of solids, including photoelectric spectroscopy.

Semiconductor physicists are familiar with the fact that while germanium and silicon are semiconductors (characterized by an energy gap between occupied and unoccupied electronic band states) grey tin, HgSe, and HgTe, belonging to the same family, are semimetals. This has been attributed[30] to the relativistic increase of the electron mass near the core of the heavy elements Sn and Hg. This increase lowers the energy of the s-like conduction states, thus closing the gap and transforming the semiconductor into a semimetal.[31,32]

The energy bands of narrow band semiconductors such as InSb are non-parabolic, i.e., their corresponding mass increases with increasing velocity.[33] This effect is similar to that postulated by Einstein[15] for relativistic free electrons. For narrow gap semiconductors, however, the electrons and holes near the gap are neither relativistic nor free. They are affected by the periodic potential of the crystal lattice. Curiously, the mass of such classical electrons has properties rather similar to those of free relativistic electrons. A simple calculation using the "$k.p$ method"[34] leads to the following "non-parabolic" expression for the energy bands near the gap (see Fig. 3):

$$E = \pm\sqrt{[(E_g/2)^2 + (\frac{\hbar}{m}P)^2 k^2]} \quad (9)$$

This equation is isomorphous to that which applies to Dirac's relativistic electron and positron, ($E_g/2$) corresponding to the rest mass of the particles. In Eq. (9) the + sign corresponds to conduction electrons and the – sign to valence holes (equivalently, electrons and positrons in Dirac's relativistic electron theory). The use of Eq. 9 has become standard in modern semiconductor physics and electronics.

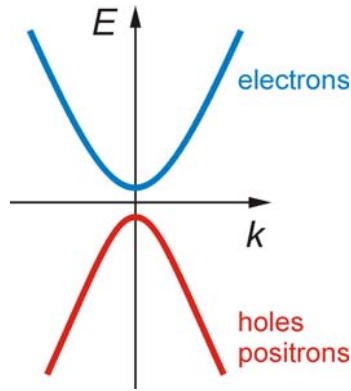

**Fig. 3** Conduction and valence bands of a narrow gap semiconductor (e.g. InSb) around a gap at $k=0$, according to Eq. 9. The upper curve represents conduction electrons (free electrons in the relativistic case) whereas the lower curve represents holes (positrons in the relativistic case).

## 2. THERMAL PROPERTIES OF SOLIDS

After the "*annus mirabilis*" Einstein spent considerable time trying to develop a microscopic theory of the thermal properties of solids, a complex which was then, and still is, central to the field of solid state physics. This was, in the early 1900s, a rather difficult task. The static crystal structures of simple materials (e.g. diamond) were becoming available but basically, nothing about their dynamical properties (e.g. phonon dispersion relations) was known. Fermi-Dirac statistics, and the details of electronic excitations in metals, only became known in 1926. Einstein's work on thermal properties thus applies to insulators although in several of his papers he contrasts his results against existing data for metals. We shall discuss here his pioneering work on the specific heat (of insulators), still relevant today, and his (by his own admission) unsuccessful attempt to develop a theory of the thermal conductivity.

### 3.1 The specific heat of insulators[35].

Einstein realized that atoms in solids vibrate around their equilibrium positions on account of their thermal energy. He assumed, for simplicity, that there was only one vibrational frequency, taking it to be an average if more than one frequency was present. This average frequency is now called the Einstein frequency $v_E$ and one speaks of the Einstein single oscillator model. All that was known at that time concerning the specific heat $C_v$ is that at high temperatures it tends asymptotically to the Petit and Dulong's value of 5.9 calories/mole K, for all substances.[36] Available data for diamond indicated a sharp drop in $C_v$ for T< 1000 K (see Fig. 4).

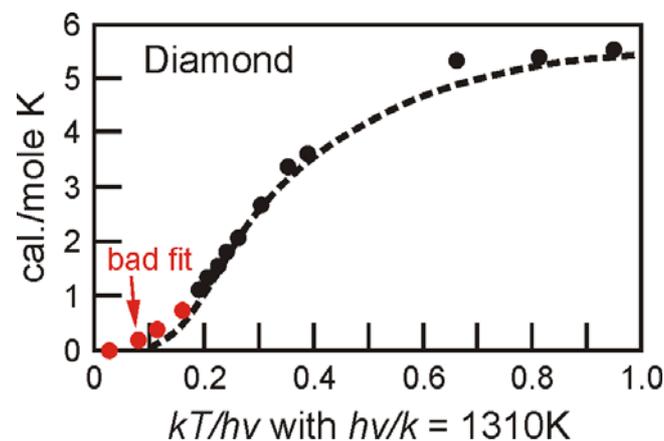

**Fig. 4.** Specific heat of diamond vs. $kT / hv_E$. The points are experimental, the dashed curve represents a fit with a single Einstein oscillator ($v_E$ = 903 cm$^{-1}$). The experimental points lie above the fitted curve for $kT / hv_E <$ 0.2. This discrepancy becomes smaller when two oscillators are used for the fit[39]. It disappears when using the Debye model[38]. From Ref. 35.

Einstein explained this behavior by assuming an ensemble of harmonic oscillators of frequency $v_E$ equal to 3 times the number of present atoms. Using Planck's ansatz for the thermal energy of one mole of these oscillators he found, for the average thermal energy, the expression:

$$<E> = \frac{3R}{N} \frac{\beta v_E}{e^{\frac{hv_E}{k_B T}} - 1} \quad (10)$$

Where $R$ is the gas constant (per mole). In present day's notation we would set $\beta = h/k_B$. By differentiating the energy $<E>$ with respect to $T$ Einstein obtained:

$$C_v = \frac{d<E>}{dT} = 5.9 \frac{e^{\frac{h\nu_E}{k_B T}} (h\nu_E / k_B T)}{(e^{\frac{h\nu_E}{k_B T}} - 1)^2} \quad , \tag{11}$$

Equation 11 yields the Petit-Dulong[36] limit for $T >> h\nu_E/k_B$. Einstein also mentioned in Ref. 35 that, at least around room temperature, the contribution of free electrons (e.g., in metals) to $C_v$ should be negligible.

Einstein fitted existing data for diamond with Eq. 11, using the Einstein frequency as an adjustable parameter which turned out to be 1310 K (in units of temperature, corresponding to 909 wavenumbers (wn) or 11.0 μm wavelength). Towards the end of Ref. 35 Einstein mentions that diamond should show infrared absorption at this wavelength, but such absorption was not known. Obviously, he knew very little about the nature of the ir absorption due to lattice vibrations, which now we know is "dipole forbidden" for the diamond structure. Raman scattering by these vibrations is, however, allowed, having been observed at 1330 wn (7.5 μm), a frequency considerably higher than the Einstein frequency (909 wn) which corresponds to an average frequency whereas, now we know, the Raman frequency is close to the maximum phonon frequency of diamond (1330 wn).[37]

The fit displayed in Fig. 4, is rather good for $h\nu/kT > 0.2$. For lower values of $T$ the values of $C_v$ obtained from the fit lie below the experimental points. This results from the assumption of a single oscillator. Five years later P. Debye introduced the elastic vibrations (Debye's) model[38] which represents a continuum of vibrational frequencies extending all the way to zero frequency. In 1911, however, Nernst and his graduate student Lindemann had already improved matters by performing a fit with two oscillators.[39] One may say "big deal"; two adjustable parameters will always give a better fit than one. However, Nernst (1920 Nobel prize for chemistry) and Lindemann (Chief scientific adviser of Churchill during WW II) were smarter than that. They used two oscillator frequencies but constrained one of them to be half the other, a fact which boiled down to using a single adjustable parameter. The two frequencies, we now know, correspond to two averages of the acoustic and optic phonon frequencies (see Fig. 5b). The Debye, Nernst-Lindemann, and Einstein fits to experimental data for diamond are shown in Fig. 5a.

In 1911[40] Einstein realized that the hypothesis of a single oscillator frequency was only a rather rough approximation. Because of the wide amplitude of the atomic vibrations in a solid, he reasoned, there should be considerable interaction between them which should transform the single frequency into bands. He then conjectured that the atomic vibrations must be strongly anharmonic. In a footnote to Ref. 40 he goes as far as to say *Our mechanics is not able to explain the small specific heats observed at low temperatures*. While writing Ref. 40, he got from Nernst the proofs of Ref. 39, with the two-frequency model. He seems to have liked it: after all, it got rid of most of the discrepancy between his model and the experimental data. He comments "…the N-L ansatz is equivalent to assuming that the atoms vibrate half of the time with a frequency ν and the other half with the frequency ν/2. The important deviation from the monochromatic behavior thus finds in this way its most primitive[41] expression". In the next paragraph he realizes that crystals must have two kinds of vibrations: acoustic and optic. In the former, he says, an atom vibrates against all neighbors whereas in the latter, a given atom vibrates against the nearest neighbors, i.e. in the opposite direction to them; not bad as a qualitative description of lattice dynamics!

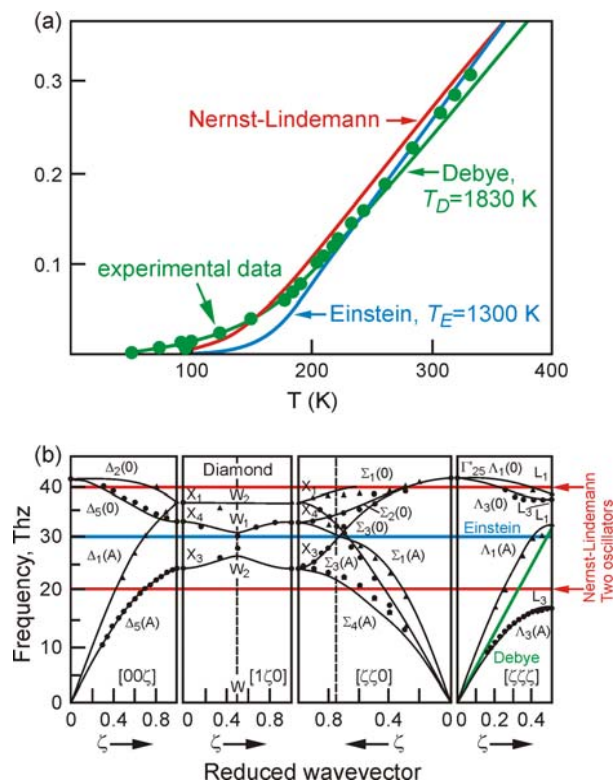

**Fig. 5(a)** The measured specific heat of diamond (dots) as compared with three fits: Einstein (blue line), Nernst-Lindemann (red line) and Debye (green line).
**5(b)** The phonon dispersion relation of diamond as compared with the two single frequencies used in the Nernst-Lindemann model. From Ref. 74.

Having taken a liking to Lindemann[42] Einstein considered the famous Lindemann's theory of melting[43] which enabled him (and before him Nernst) to derive average vibrational frequencies from the crystal's melting temperature. He is then pleasantly surprised by the good agreement of these frequencies with those obtained through fits of the temperature dependence of $C_v$. He expresses some displeasure at the fact that the Lindemann frequency agrees better with the specific heat vs. *T* than the frequencies he obtained by comparing the optical vibrations with the bulk modulus.[44] Because of its simplicity, I shall spend a few words on the method used in Ref. 44 to relate the bulk modulus to the "Einstein frequency", as illustrated in Fig. 6.

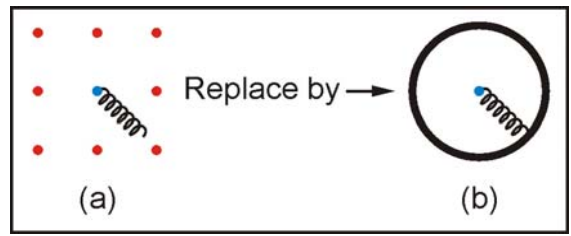

**Fig. 6(a).** Schematic diagram of a fictitious crystal used by Einstein[44] in order to derive a relationship between the bulk modulus and the Einstein frequency. (b) Similar to (a), but symmetrized in order to simplify the calculation.[44]

Einstein represents a solid as a periodic array (Fig. 6a). He then connects the central atom with its nearest neighbors by equal springs. The force constant of the springs can be determined from the bulk modulus, which corresponds to a uniform compression of Fig. 6(a). Once the force constant is known, it is trivial to obtain the frequency of vibration of the central atom against the surrounding ones. Somehow Einstein seems to have regarded the summation over all 8 nearest neighbors as too menial and tedious, so he replaced the peripherical atoms by a sphere, smearing their masses uniformly over the sphere. It is then trivial to obtain the relationship between the "Einstein" frequency and the bulk modulus. Einstein mentions having picked up this idea from a paper by Sutherland[45]. After Ref. 44 appeared, Einstein realized that already Madelung had derived a quantitative relationship between elastic constants and the "Einstein" frequency[46] For reasons unbeknownst to me, Einstein assigns the priority[47] for the discovery of what he calls " this fundamental and important relation between the elastic and the optical behavior of solids". Reference 47 appeared in the *Annalen* as a regular article; nowadays it would be simply a comment or an erratum. One should mention at this point that Einstein's publications list contains many such short articles correcting errata, priorities, or presenting complementary aspects which had been omitted in the main articles.

3.2 **Thermal conductivity** [40]

After having successfully tackled the problem of the specific heat of insulators, Einstein tries to develop a theory of heat transport, i.e. of the thermal conductivity $\kappa\ (T)$. This is a much more complex problem and too many building blocks were missing at the time. Even now, first principles calculations of the thermal conductivity of simple solids are rather incomplete.[48] Einstein assumed that the heat transport takes place through the interaction between a thermally excited atom and its nearest neighbor down the temperature gradient (Fig. 7). He had attributed the width of the vibrational frequency band to this coupling so he now estimated the coupling from the conjectured bandwidth. He then derives an expression for $\kappa\ (T)$ which is proportional to the specific heat. Using

Petit and Dulong's value[36] for the latter, Einstein reaches the conclusion that the thermal conductivity should be, at room temperature, much smaller than the measured one. It should also be, in the Petit and Dulong region, independent of $T$, contrary to the decrease with $T$ that had been experimentally observed. He concludes with one of his typical statements: *We must thus conclude that mechanics is not able to explain the thermal conductivity of insulators. Moreover, the assumption of a quantized energy distribution does not help…*

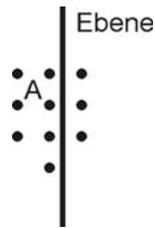

**Fig. 7** Diagram used by Einstein in order to illustrate the origin of the thermal conductivity. *Ebene*, which means "plane" in German, represents a plane that separates the hotter region from the colder region .[40]

After using some rather scurrilous dimensional argument he is able to derive the $\kappa \sim T^{-1}$ law proposed by Eucken but, surprisingly, concludes with the statement:
*The task of the theory will be to modify molecular mechanics in such a way that it can account for the specific heat as well as the apparently so simple laws governing the thermal conductivity*

We now know that the laws governing the thermal conductivity and its dependence on temperature and *isotopic mass* are not so simple.[48,49] A curve illustrating the standard behavior of the thermal conductivity of an insulator vs. temperature $\kappa\,(T)$ is shown in Fig. 8. The curve shows three distinct regions, one at low $T$, proportional to $T^3$, the high temperature region, in which $\kappa\,(T)$ decreases rapidly with increasing $T$, and a maximum which can be varied by changing the isotopic composition of the crystal (isotopes were unknown to Einstein in 1907!).

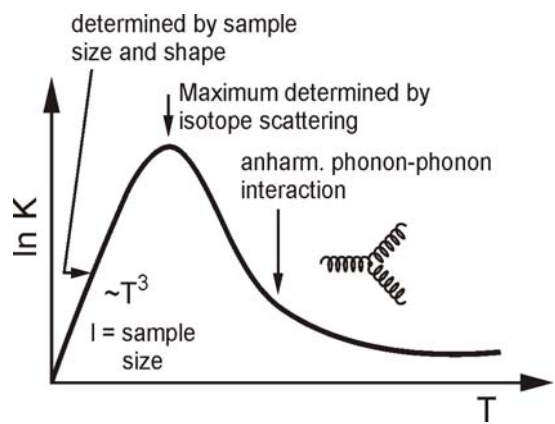

**Fig. 8.** Schematic diagram of the thermal conductivity of an insulator. The low-$T$ part ($\sim T^3$) corresponds to ballistic phonon transport, whereas the high temperature part corresponds to anharmonic processes, mostly of Umklapp-type.[49]

In order to describe the effects leading to these three regions, we write the thermal conductivity as:

$$\kappa(T) = (1/3)\, C_v \cdot v \cdot l \;, \tag{12}$$

where $v$ represents an average velocity of the acoustic phonons and $l$ their mean free path. In the low temperature region, $l$ would be larger than the sample dimensions (ballistic heat transport). It thus becomes of the order of those dimensions and temperature independent. The $T^3$ law is obtained by considering that $v$ and $l$ are independent of temperature whereas the specific heat $C_v$ is, according to Debye, proportional to $T^3$. At high temperatures the mean free path of the phonons which transmit the heat decreases rapidly with increasing $T$ because these phonons collide with thermally excited phonons through anharmonic interactions. The maximum between these two regions is due to phonon scattering by the fluctuation of atomic masses resulting from the presence of different (stable) isotopes. All these processes are certainly not as simple as Einstein had envisaged them. A few decades had to elapse before all theoretical ingredients required to explain $\kappa(T)$ became available. Having realized the difficulties involved, Einstein moved to greener pastures and left the theory of thermal conductivity to future generations.

4. **THE BOSE-EINSTEIN STATISTICS**

**4.1.1   The zero-point energy**

In an article coauthored with Otto Stern (1943 Nobel laureate in physics).[50] Einstein proposed a rather ingenious way of deriving the zero-point energy of an oscillator. This proposal is particularly remarkable: we now believe the zero point motion to be a consequence of quantum mechanical uncertainty which was totally unknown in 1913. When it was established in 1924, Einstein became very skeptical about it (*God does not play dice!*) in spite of the fact that he had introduced the concept and made early use of it.

Einstein and Stern reasoned as follows: By expanding Eq. 10 in the high temperature limit we find, for a single one-dimensional harmonic oscillator:

$$E = \frac{h\nu}{e^{\frac{h\nu}{k_B}} - 1} \to k_B T - \frac{h\nu}{2} \;. \tag{13}$$

When comparing Eq. 13 with the result obtained from classical statistics $E \to k_B T$, which should be valid at high temperatures, they were disturbed by the presence of the negative energy $-h\nu/2$. So they added to the r.h.s. of Eq. 13 the term $+h\nu/2$, in an *ad hoc* manner (see Fig. 9, copied verbatim from Ref. 50 so as to illustrate the kind of figures used those days, figures are rare in Einstein's papers anyhow). For $T \to 0$, Eq. 13 then remained finite and equal to $+h\nu/2$. They interpreted this fact as signaling the existence of "thermal" motion in a harmonic oscillator even for $T \to 0$, ten years before Heisenberg derived this result from his celebrated uncertainty principle.[51]

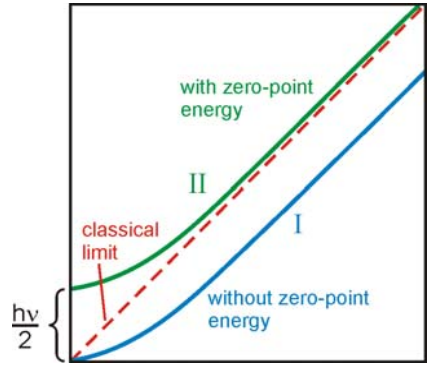

**Fig. 9** Diagram used by Einstein and Stern to illustrate the need of a zero-point vibrational energy ($h\nu_E$) in order to bring the Bose-Einstein distribution to agree with the classical one in the high temperature limit (dashed line). [50]

Einstein (and Stern) as usual, tried to find some experimental verification of the zero-point energy.[50,51] They had come across recent data on the specific heat of hydrogen at low temperature which they thought was due to the energy accumulated in a rigid rotator (the $H_2$ molecule). Knowing nothing about quantum mechanics, they equated the rotational energy to the vibrational energy of a harmonic oscillator with vibrational frequency equal to the rotational frequency of $H_2$, with and without zero-point energy. They used these relationships to determine $v$, and the corresponding energy, vs. $T$. The specific heat vs. $T$ obtained by differentiating the so-obtained rotational energy vs. T agreed better if the zero-point energy was added to the corresponding harmonic oscillator:

$$E = \frac{J}{2}(2\pi v)^2 = \frac{h\nu}{e^{\frac{h\nu}{k_B T}} - 1} + \frac{h\nu}{2} \qquad (14)$$

We now know that this procedure is not correct. The quantum rigid rotator is not equivalent to a harmonic oscillator; among other differences it has no zero-point energy.[52]

The experimental $C_v$ (T) is now known to be strongly affected by spin statistic and transitions from ortho- to para-hydrogen. The existence of spin, and the corresponding two modifications of $H_2$, was of course unknown to Einstein and Stern: Eq. 14 has to be regarded as another "act of desperation", in this case an unsuccessful one. Nevertheless, the incorrect hypothesis represented by Eq. 14 has generated considerable literature (72 citations), especially in recent years. For a detailed discussion, see Ref. 52. In spite of the shortcomings of Ref. 50 just discussed, the correct derivation of the zero point energy of an oscillator, without prior knowledge of the uncertainty principle, is certainly an admirable *tour de force*.[52]

4.2 **The quantum theory of radiation**[53]

Until 1907 Einstein published all his articles in the "Annalen". In 1907 he begins to diversify[54] using the "Annalen" less and less and, increasingly, the *Proceedings of the Royal Prussian Academy of Sciences*, of which he became a member in 1913, with 21 positive votes and a negative one (see Ref. 9). From 1908 on he also used the *Physikalische Zeitschrift* as a medium: I am not aware of the reason for his moving away from the "Annalen". Some of the work he published in the "Zeitschrift" is of a more applied nature, including an article in which he proposes the use of Zn and Cd (two metals with a rather small work function, $I$ ~4.1 eV) as photocathodes for ultraviolet photometry.[55] In 1917 he published in the "Zeitschrift" a much-celebrated paper under the title "The quantum theory of radiation"[56]. Quantum theory was slowly approaching but had not yet arrived. In this paper Einstein used semi-classical arguments to develop some of the most important concepts and equations of the quantum theory of radiation. Einstein realized that atoms or molecules are excited in the presence of radiation of the "right phase" and frequency. Such systems, if excited, can be de-excited under the presence of radiation of the "wrong phase". These two processes of excitation and de-excitation are now known as absorption and stimulated emission of light. Atoms and molecules can also be de-excited spontaneously, without the presence of external radiation, thus leading to the concept of spontaneous emission. We now know that spontaneous emission is effected by the zero-point electromagnetic energy, a concept not available in 1917. For the radiation of frequency $v$ to be in thermal equilibrium the absorption processes must equal the sum of the two types of emission processes and Planck's black body distribution must hold.

Einstein describes the strength of the absorption and the stimulated emission by a coefficient labeled $B$ which he finds to be the same for both types of processes. He represents the spontaneous emission by a coefficient A which he finds to be related to B though the famous *Einstein relation* (one of many relations that bear his name. See e.g., Eq. 8):

$$A = \frac{8\pi h v^3}{c^3} B \tag{15}$$

Lasers, invented several decades later, are based on the phenomenon of stimulated emission first predicted by Einstein.

4.3 **Einstein, Bose, and Bose –Einstein statistics.**

On June 4[th], 1924 Satyendra Nath Bose, a 30 years old reader (associate professor) at the University of Dacca (then India, now Bangladesh) sent Einstein a manuscript in English with a covering letter full of praise, asking him, in no uncertain terms, to translate it into German and submit it for publication to the "Zeitschrift für Physik". Rumor has it that the manuscript had been rejected previously by the Philosophical Magazine. Einstein was very pleased by the manuscript and proceeded to do as requested. The paper appeared in print on July 7, 1924. The logistics of this case puzzles me a bit: the manuscript was sent to Einstein from Dacca, a provincial town (I presume by mail, the only rather dubious alternative would have been cable) and it appeared in print one month later after having been translated by Einstein himself. Modern day editors, take heed! Bose's work derives the Planck (from now one called Bose or Bose-Einstein) distribution without superfluous interactions with additional particles or radiation, making simply use of statistics and the assumption of the indistinguishability of particles: pairs of particles AB and BA count as a single state, not two. At the end of the printed article[57], and having identified himself as a translator, not a coauthor, Einstein added:

Note of the translator: *Bose's derivation of Planck's formula represents, in my opinion, real progress. The method used in this paper also can be used to derive the quantum theory of ideal gases, as I shall show elsewhere.*

This is probably the most encomiastic praise Einstein ever wrote concerning the work of a colleague.

So far so good. Encouraged· by his success, Bose sent within a short time, a second paper to Einstein which the latter also translated and submitted for publication to the *Zeitschrift für Physik*. The logistics here is even more puzzling. The publication[58] bears the date of receipt of July 7, 1924 and the presumable date of mailing of June 14, 1924. Blanpied,[59] however, claims that this second paper was sent to Einstein on Oct. 26, 1924, which would make much more sense, having given Bose time to have heard about the acceptance, let alone publication, of the first paper. Unfortunately, it seems to have occurred to Bose that there should be no such thing as "stimulated emission of radiation", an idea that he buttressed with erroneous algebra. No stimulated emission would have meant no lasers! In the published paper, Einstein, without identifying himself as the translator, added a signed note (a full page!) blasting at Bose and showing that there must indeed be stimulated emission, by means of two simple arguments. The first one is especially simple and appealing: in the classical theory of interaction of a resonant dipole with electromagnetic radiation, absorption as well as stimulated emission appears on the same footing. Depending on the phase of the radiation, absorption takes place. For the opposite phase emission of radiation occurs. Einstein points out that the classical theory is simply a limiting case of the quantum theory and therefore the absence of stimulated emission proposed by Bose must be wrong. It may sound strange that Einstein would translate this paper and endorse it for publication, while adding a note to the printed article saying, in no uncertain terms, that it was wrong. It seems that he also wrote a letter to Bose mentioning the pitfalls of the work and that Bose answered that he was preparing a manuscript in which Einstein's objections would be dispelled. This correspondence, however, is not extant.[59] The manuscript, if ever written, did not appear in print.

In spite of his long life (1894-1974, he even lived to see the advent of the laser), Bose hardly published anything of relevance after his groundbreaking 1924 paper. He is highly revered in India and most Indian biographical material simply glosses over the existence of the "second paper".[59,60]

Upon reading Ref. 57, it must have dawned upon Einstein that the Bose-(Einstein) distribution law applied to massless particles (photons, vibrons, phonons) whose number increases with increasing temperature. He then generalized it to massive particles (of the type now called Bosons) whose number is conserved (The difference between Bosons and Fermions was not known to Einstein at the time). In three articles he presented at separate meetings of the Prussian Academy [61,62,63] he describes the generalization of Bose's derivation to apply to massive particles. In spite of the irritation which must have caused Bose's "second paper"[58] he gives ample credit to the latter for having derived Planck's formula on the basis of the indistinguishability of particles. He points out that, by means of this assumption, one is able to rescue "Nernst's theorem" (the third principle of thermodynamics): at $T=0$ there is only one state if the particles are indistinguishable and therefore the entropy vanishes. He ends Ref. 61 in a typical Einstein way, mentioning a paradox "which he has been unsuccessful in solving". He considers two slightly different kinds of molecules. Since they are distinguishable, their statistical behavior will be different than if the molecules were the same (i.e. indistinguishable). He then expresses his difficulties in understanding the discontinuous transition from a set of equivalent molecules to two sets of slightly different ones.

Einstein's distribution function for a set of $N$ equivalent massive molecules is:

$$f(E) = \frac{z(E)}{e^{\frac{E-\alpha}{k_B T}} - 1} \qquad (16)$$

Where $z(E)dE$ is the number of states with energy between $E$ and $E + dE$. The "chemical potential" $\alpha$ is determined from the condition:

$$\int_0^\infty f(E)dE = N \qquad (17)$$

Einstein realized that for $T=0$ all molecules are in the lowest energy state and, *because of indistinguishability*, they correspond to only one statistical state. As $T$ increases, molecules begin to evaporate from this state and to occupy a range of energies, whereas the lowest energy state remains multiply occupied up to a temperature $T_{BE}$. This experimentally somewhat elusive phenomenon is called Bose or Bose-Einstein condensation (although the idea occurred to Einstein alone, after reading Ref.57). In Ref.62 he suggested as possible candidates for the observation of the BE condensation $H_2$ and $^4$He. The condensation should appear as a sharp decrease in the viscosity. He even used the term "*superfluidity*" and estimates that $T_{BE}$ should be about 40 K.

Superfluidity in $^4$He ($T_S$ = 2.17 K) was discovered by Kapitza in 1937[64] (Nobel laureate, 1978). Tisza suggested in 1938 that superfluid $^4$He is a Bose-Einstein condensed gas[65], but it has been pointed out that $^4$He is not an ideal gas: the $^4$He atoms interact strongly.[66] For a detailed discussion of the modern theory of $^4$He see [67]. Since the 1960s there has been considerable activity trying to prove the presence of Bose-Einstein condensation in insulators.[68] The particles that should condense are either excitons or polaritons. Among the most investigated crystals are CuCl and $Cu_2O$. Although some evidence of the formation of a coherent condensate of these particles, similar to a Bose-Einstein condensate, has been obtained, this evidence is not yet conclusive. Conclusive evidence was obtained ten years ago for highly diluted rubidium vapor at extremely low temperatures.[69]

## 5. SUPERCONDUCTIVITY

We have seen above that Einstein tackled usually, but not always successfully, almost all important problems of condensed matter physics. He even dealt with superfluidity. It is therefore surprising that no Einstein publication concerning superconductivity appeared in the "standard" literature. He felt tempted to get involved in 1922, on the occasion of the 40$^{th}$ anniversary of H. Kamerlingh Onnes having become a professor at Leiden (he had discovered superconductivity in 1911 and received the Nobel Prize in 1913). Einstein wrote an article on the theory of superconductivity which was published in the Kamerling Onnes's 40$^{th}$ anniversary Festschrift.[70] At that time Fermi statistics was not known but this did not deter Einstein. He noticed that even in normal metals the electrical resistance should vanish for $T \to 0$ but this is not the case. He gives credit to K.O. for having realized that the residual resistance depends strongly on residual impurities. He presents some interesting considerations on the nature of electrical conduction revealing, through a glass darkly, the phenomenon of band conduction and Mott transitions. He concludes that superconductivity must be related to the existence of a coherent state connecting the outer electrons of an atom with those of its neighbors. He then postulates that impurities must destroy that coherence, in particular when a foreign atom interrupts a superconducting chain.

Again typical of Einstein, he mentions, in a *note in proof* that the aforementioned speculations have been laid to rest by a recent experiment of Kamerlingh Onnes who has shown that a junction between two superconducting metals (lead and tin) also exhibits zero resistance, i.e. superconductivity.

## 6. RECENT APPLICATIONS OF EINSTEIN'S EARLY WORK

### 6.1 The diffusion-mobility relation

It has been mentioned in Sect. 2.3 that the diffusion-mobility relation [Eq. 8] has become rather important in the realm of semiconductor technology. There have therefore been recent efforts to generalize it to cases not covered by Einstein's original considerations.[71,72] Equation 8 was derived under the assumption of Boltzmann statistics, which applies to lightly doped (non-degenerate) semiconductors. The simplest generalization concerns the use of Fermi statistics and, in particular, its degenerate limit.

Band non-parabolicity (Fig. 3) can also be important. Appropriately generalized expressions are given in Table 1 of Ref. 71. Very recently, a publication with generalizations of the Einstein relation to lower dimensional systems and nanostructures has appeared.[72]

### 6.2 The Einstein oscillator model for the temperature dependence of physical properties on temperature [73,74]

We have discussed in Sec.3.1 the use of a Bose-Einstein term, involving an average frequency, to represent the temperature dependence of the specific heat of insulators, the so-called Einstein oscillator model. We have also mentioned the generalization to two oscillators[39] and the Debye ansatz to describe the specific heat at very low temperatures ($C_v \sim T^3$). Similar terms have been used to describe the temperature dependence of other physical properties such as the elastic constants, the phonon frequencies, the thermal expansion, the optical energy gaps, etc. Einstein fits have been used to obtain the renormalization of such properties by the thermal agitation at $T \to 0$.[73,74] I display in Fig. 10 the measured temperature dependence of the indirect exciton frequency of diamond, together with a single oscillator fit with the fitted Einstein frequency of 1080 wn (~1580 K). From this fit, the zero-point exciton renormalization of 370 meV is obtained. This number has been used to estimate the hole-phonon interaction which turns out to be rather large as compared with that in Ge and Si.[74] It has been suggested[74] that this large hole-phonon interaction is responsible for the superconductivity recently observed in boron-doped diamond, with a critical temperature close to 10 K.[75,76]

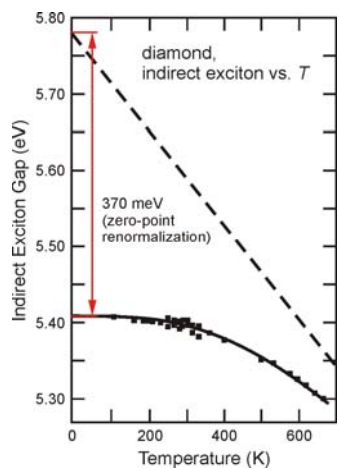

**Fig. 10** Temperature dependence of the photoemission at the indirect exciton energy of diamond. The points are experimental, the solid line a fit with a single Einstein oscillator. The dashed line represents the classical result extrapolated to $T = 0$. This extrapolation yields the zero point gap renormalization (370 meV). From Ref.74.

## 7. CONCLUSIONS

When I started working on this manuscript, I had in mind reading a few of Einstein's publications (in the German original, of course) pertaining to what we now call Solid State or Condensed Matter Physics. It soon became clear to me that this work would have to cover much more ground than I originally had in mind. Correspondingly, the power point presentation would go well over one hour, the typical limit granted for an Einstein talk in this *Annus Mirabilis*. It has turned out to be a fascinating task, not only from the physics point of view, but because of providing me with interesting insights into the way Einstein worked, thought and chose his problems. I found that some results I rederived recently, had already been derived by Einstein nearly 100 years ago in exactly the same manner (e.g., the zero-point energy of the harmonic oscillator, Sect 4.1). I believe that the reader will agree with me that Einstein well deserves to be called the father of Condensed Matter Physics.

## ACKNOWLEDGMENTS

I would like to thank Karl Syassen for a critical reading of the manuscript and the Fond der Chemischen Industrie for financial support.

# References

[1] Consequences of the observations of capilarity phenomena, Ann. d. Physik **4**, 513 (1901).

[2] Calculation of surface enthalpy of solids from an ab initio electronegativity based model: case of ice, J.M. Douillard and M. Henry, J. Colloid and Interface Science **263**, 554 (2003).

[3] On the thermodynamic theory of the potential difference between metals and fully dissociated solutions of their salts and on and electrical method to investigate molecular forces, Ann. d. Physik Ann. d. Physik **8**, 798 (1902).

[4] See, for instance: Theory of the opalescence of homogeneous fluids and mixtures of fluids near the critical state, Ann. d. Physik, **33**, 1275 (1910). Cited 660 times.

[5] Kinetic theory of thermal equilibrium and the second principle of thermodynamics, Ann. d. Physik **9**, 417 (1902).

[6] A theory of the foundations of thermodynamics, Ann. d. Physik **11,** 170 (1903). Cited 42 times.

[7] Concerning the general molecular theory of heat, Ann. d. Physik **14**, 354 (1904). Cited 33 times.

[8] M. v. Smoluchowski, Ann. d. Physik **25**, 205 (1908).

[9] See E. Pais, Subtle is the Lord… (Oxford Univ. Press, Oxford, 1982), p.55.

[10] A. Einstein, Ann. d. Physik **34**, 175 (1911).

[11] A new determination of the molecular dimensions, Ann. d. Physik **19**, 289 (1906) Cited 1622 times!

[12] On the production and transformation of light according to a heuristic point of view. Ann. d. Physik **17**, 132 (1905), cited 401 times.

[13] On the motion of particles suspended in a liquid at rest, as required by the molecular-kinetic theory of heat. Ann. d. Physik **17**, 549 (1905), cited 1520 times.

[14] On the electrodynamics of moving bodies Ann. d. Physik **17**, 891 (1905) cited 720 times

[15] Does the inertial mass of a body depend on its energy content? Ann. d. Physik **18**, 639 (1905). Cited 106 times

[16] On Dec. 10, 1922, the day of the presentation of the Nobel Prizes, Einstein was traveling in the Far East. The prize was received on his behalf by the German ambassador to Sweden.

[17] G.N. Lewis, Nature **118**, 874 (1926).

[18] Einstein used in Ref. 9 Lenard's data on photoemission and Stark's data on the ionization of gases by uv light. Lenard was awarded the Nobel Prize for 1905 whereas Stark received the award for 1919. Through the vagaries and ironies of history both Lenard and Stark became the leading representatives of the Aryan Physics movement which accompanied the groundswell of Nazi sentiment in Germany. As an example I quote from a newspaper article coauthored by both physicists in 1924, well before de Nazi access to power: "…a racially alien spirit has been busily at work for over 2000 years. The exact same force is at work, always with the same Asian people behind it that had brought Christ to the cross, Jordanus Brunus to the stake…" See K. Hentschel, Physics and National Socialism (Birkhäuser Verlag, Basel, 1996) p.7.

[19] Z.X. Shen et al., Science **267**, 343 (1995).

[20] J. Schäfer et al., Rev. Sci. Inst. 64, 653 (1993).

[21] C. Sebenne et al., Phys.Rev.B 12, 3280 (1975).

[22] Actually one should say that Ref. 11 was based on his thesis which was submitted to the University of Zürich on April 30. 1905. Ref. 11 did not appear till 1906.

[23] On the theory of Brownian motion, Ann. d. Physik, **19**, 371 (1906), cited 496 times.

[24] M. Gouy, J. de Physique, **7**, 561 (1988).

[25] Can quantum-mechanical description of reality be considered complete? A.Einstein, B. Podolsky, and N. Rosen, Phys. Rev. **47,** 777 (1935). Cited 2409 times.

[26] A comment on Ref.9. Ann. d. Physik **34**, 591 (1911). Cited 993 times.

[27] P.A.M. Dirac, Proc. Roy. Soc. London, Ser. **A 123**, 714 (1929). Here Dirac states that relativistic effects do not affect the atomic and molecular structure. If this were correct, similar conclusions would apply to solids.

[28] Relativistic effects in the properties of gold, P.Schwerdtfeder, Heteroatom Chemistry **13**, 578 (2002)

[29] On the classical radiation of accelerated electrons, J.Schwinger, Phys.Rev. **75**, 1912 (1949). Cited 650 times

[30] F. Herman, in Proc. Int. Conf on the Physics of Semiconductors (Dunod, Paris, 1964) p.2-22.

[31] P.Y. Yu and M. Cardona, Fundamentals of Semiconductors, third edition (Springer, Heidelberg, 2005) p.95.

[32] Relativity and the periodic system of the elements, P.Pyykkö and J. P. Desclaux, Accounts of Chemical Research, **12**, 276 (1979); Why is mercury liquid? – Or, why do relativistic effects not get into chemistry textbooks? L.J. Norrby, J. Chem. Education, **68**, 110 (1991).

[33] W.G. Spitzer and H.Y. Fan, Phys. Rev. **106**, 882 (1957).

[34] Ref. 31, p.105.

[35] Planck's theory of radiation and the theory of the specific heat, Ann. d. Physik **22**, 180 (1907). Cited 300 times.

[36] A.T. Petit and P.L. Dulong, Ann. Chim. Phys. **10** 395 (1819).

[37] C. Ramaswamy, Indian J. Phys. **5**, 97 (1930); Nature **125**, 704 (1930).

[38] P. Debye, Ann. d. Physik **39**, 789 (1912). Cited 453 times.

[39] W. Nernst and F.A. Lindemann, Z. Elektrochemie **17**, 817 (1911). Cited 131 times.

[40] Elementary considerations about thermal motions of molecules in solids, Ann. d. Physik 35, 679 (1911), 82 citations.

[41] The word „primitiv" in German does not have exactly the same meaning as in English. It probably would be better translated as "simple". Not knowing exactly what Einstein had in mind, I left the word "primitive" in the translation.

[42] F.A. Lindemann, Viscount Cherwell, an Englishman of German ancestry, was the thesis adviser of R.V. Jones, who was the thesis adviser of W. Paul, my thesis adviser.

[43] F.A. Lindemann, Physik. Zeitschr. **11**, 609 (1910), cited 870 times.

[44] A relation between the bulk modulus and the specific heat in monatomic solids, Ann. d. Physik **34**, 170 (1911). Cited 95 times.

[45] W. Sutherland, Phil.Mag. **20**, 657 (1910).

[46] E. Madelung, Physik. Zeitschr. **11**, 898 (1910).

[47] Comments on my work: "A relation between the elastic behavior…" Ann. d. Physik **34**, 590 (1911). Cited 24 times, some of the citations being very recent. Not bad for a paper whose only purpose is to set priorities straight.

[48] A. Sparavigna, Phys. Rev. B **67**, 144305 (2003).

[49] M. Asen-Palmer et al., Phys. Rev. B **56**, 9431 (1997).

[50] Some arguments in favor of the existence of molecular agitation at zero temperature, Ann. d. Physik 40, 551 (1913). Cited 72 times. This is one of the very few coauthored papers of Einstein.

[51] The existence of a zero-point energy had also been conjectured by Planck. See M. Planck, Ann. d. Physik **37**, 642 (1912).Cited 32 times.

[52] J.P. Dahl, J. Chem. Phys. **109**, 10688 (1998).

[53] The quantum theory of radiation, Physik. Zeitschrift **18**,121 (1917). Cited 720 times.

[54] Theoretical remarks on the Brownian motion, Zeitsch. für Elektrochemie, **13**, 41 (1907), cited 28 times.

[55] The use of photoelectric cadmium and zinc cells for ultraviolet sunlight photometry, Physik. Zeitsch.**15**, 176 (1914). Cited 10 times. I had some difficulties in obtaining a copy of this article. When it arrived I was dismayed to see that there was a serious error in the WoS data bank. This publication of Einstein has nothing to do with photocathodes (too bad!). Its correct title is "Basic remarks on general relativity and the theory of gravitation" (makes more sense, doesn't it?). The article on photocathodes actually exists. It was authored by Elster and Geidel and appeared in Physik. Z. **15**, 1 (1914). I have alerted the ISI and hope that, in due course, the error will be corrected.

[56] A. Einstein, Physik. Zeitsch. **18**, 121 (1917). Cited 720 times.

[57] Planck's law and the hypothesis of light quanta, S.N. Bose, Z. Physik, **26**, 178 (1924). Cited 210 times.

[58] Thermal equilibrium in the radiation field in the presence of matter, S.N. Bose Zeitsch. für Physik **27**, 384 (1924). Cited 16 times.

[59] Satyendranath Bose: co-founder of quantum statistics, W.A. Blanpied, Am. J. Phys **40**, 1212(1972).

[60] See, however, Theimer and Ram, Am. J. Phys. **45**, 242 (1977).

[61] Quantum theory of the monatomic ideal gas, Sitzber. Preuss. Acad. **2**, 261 (1924). Cited ~ 150 times.

[62] Quantum theory of the monatomic ideal gas II, Sitzber. Preuss. Acad. **3**,1925. Cited ~ 100 times.

[63] On the quantum theory of the ideal gas, Sitzber. Preuss. Acad. **18**, 1925. Cited ~60 times.

[64] P.L. Kapitza, Nature **141**, 74 (1937). Cited 107 times.

[65] L. Tisza, Nature **141**, 913 (1938).

[66] L. Landau, Phys. Rev. **60**, 356 (1941).

[67] G.D. Mahan, *Many-Particle Physics* (Plenum Press, N.Y., 1990).

[68] D. Snoke, Science **298**, 1368 (2002).

[69] M.R. Andrews et al., Science **273**, 84 (1996). Cites 2421 times!. Two of the coauthors, C.A, Wieman and E.A. Cornell were awarded for this work, together with W. Ketterle, the Nobel Prize in Physics for 2001.

[70] A. Einstein, theoretical considerations concerning superconductivity in metals (in German), Festschrift in honor of H. Kamerlingh Onnes on the occasion of the 40$^{th}$ anniversary of his professorship (Eduard Ijdo, Leiden, 1922) p.429. Cited 3 times.

[71] P.T. Landsberg, Einstein and statistical thermodynamics III: the diffusion-mobility relation in semiconductors, Eur. J. Phys. **2**, 213 (1981).

[72] L.J. Singh et al., The Einstein relation in nanostructured materials: simplified theory and suggestion for experimental determination. Electrical Engineering, **87**, 1 (2005).

[73] M. Cardona, phys. stat. sol. (a) **188**, 1209 (2001).

[74] M. Cardona, Solid State Commun., **133**, 3 (2005).

[75] E.A. Ekimov et al., Nature **428**, 542 (2004).

[76] Y. Takano, Private communication.